# The surface-forming energy release rate versus the local energy release rate

Si Xiao, He-ling Wang, Bin Liu[*], Keh-Chih Hwang

*AML, CNMM, Department of Engineering Mechanics, Tsinghua University, Beijing 100084, China*

[*]Corresponding authors. Tel.: 86-10-62786194; fax: 86-10-62781824

E-mail address: liubin@tsinghua.edu.cn (Bin Liu);

## Abstract

This paper identifies two ways to extract the energy (or power) flowing into a crack tip during propagation based on the power balance of areas enclosed by a stationary contour and a comoving contour. It is very interesting to find a contradiction that two corresponding energy release rates (ERRs), a surface-forming ERR and a local ERR, are different when stress singularity exists at a crack tip. Besides a rigorous mathematical interpretation, we deduce that the stress singularity leads to an accompanying kinetic energy at the crack tip. The local ERR $G_L$ represents the driving force to overcome the surface energy and the accompanying kinetic energy, while the surface-forming ERR $G_s$ represents the driving force to overcome the surface energy only. Their advantages and disadvantages are discussed. We recommend using the surface-forming ERR $G_s$ based fracture criterion for a crack propagation in elastic-plastic materials, since it has a wide applicability and concise formulae which are easy to compute among all energy based criteria.

## 1 Introduction

For elastic-plastic fracture problems, the *J*-integral (Rice et al., 1968) based fracture criterion is widely used when there is no crack propagation. However, if a crack propagates, the plastic unloading will appear and then the strain energy density in the *J*-integral cannot be defined unambiguously. Many researchers tried to put forward different methods to solve this problem (Wnuk and Read, 1986; Roos and Eisele, 1988; Schmitt and Kienzler, 1989; Cotterell and Atkins, 1996; Thomason, 1990; Zhu, 2009; Carka et al., 2012). For example, Brust and his collaborators (Brust et al., 1985; Brust et al., 1986; Brust and Atluri, 1986) defined a path-independent integral, i.e. $T^*$ integral, by introducing the total accumulated increments of stress working density for an incremental plasticity theory. They further found that the *J*-integral and $T^*$ curves were almost coincident for a small amount of crack growth, but deviated from each other as the crack further grows. When the crack growth reaches to steady state, the *J*-integral unreasonably continues to rise, while the $T^*$ turns to be a constant. By invoking the second law of thermodynamics, Simha et al. (2008) derived the near-tip and far-field *J*-integrals for a growing crack in finite deformation regime with incremental plasticity. Although these works are very important steps to understand or solve the fracture problems of elastic-plastic materials, to the best of our knowledge, they have not been widely used yet.

We note that the energy definition is usually controversial and inconsistent among many criteria and leads to confusion for users, such as the stress working density used by Brust (Brust et al., 1985; Brust et al., 1986; Brust and Atluri, 1986), but Helmholtz free energy used by Simha et al. (2008). Therefore, adopting the power balance to avoid any energy definition should be a better starting point to study the elastic-plastic crack propagation problems. In our previous paper (Xiao et al., 2015), a surface-forming energy release rate $G_s$ is defined based on the power balance, which represents the energy available for separating the crack surfaces during the crack propagation and excludes the loading-mode-dependent plastic dissipation. We also proposed the corresponding fracture criterion, which has no limitation on the constitutive behaviors of materials and has a wider applicability. Moreover, a reasonable interpretation of Rice paradox on crack propagation in elastic-perfectly plastic materials was given.

However, an interesting contradiction in that paper was pointed out to us by Prof. Landis from the University of Texas at Austin. We find that our further investigation on this contradiction can disclose a derivation error, which has been ignored not only in our previous paper (???), but also in some textbooks and lecture notes (Bower, 2005; Suo, 2016). This investigation thus can deepen our understanding on the fracture mechanisms and behaviors.

The paper is structured as follows. In Section 2, we will introduce two energy release rates based on the power balance of the area within a contour, and then point out an interesting contradiction between them. Concise formulae of the two energy release rates are derived first and their physical meanings are illustrated and compared in Section 3. In Section 4, we discuss several issues on determining and simulating crack propagation in elastic-plastic materials. The conclusions are summarized in Section 5.

## 2 An interesting contradiction between two energy release rates

As the energy cannot be defined unambiguously, the energy release rate can be introduced through the power balance during the crack propagation. There are two ways to establish the power balance relations. The first one is to investigate the power balance within a fixed contour $\Gamma$ surrounding the crack tip as shown in Fig. 1. $x_1, x_2$ is a stationary coordinate system (fixed on the material points), $A$ is the area enclosed by the contour, and $\mathbf{n}$ is the unit external normal vector. The power balance during a crack propagation can be written as

$$G_s \dot{a} = \int_\Gamma n_j \sigma_{ij} \dot{u}_i d\Gamma - \int_A \sigma_{ij} \dot{\varepsilon}_{ij} dA \tag{1}$$

where $G_s$ represents the power available for separating the crack surfaces and is named as the surface-forming energy release rate (ERR). $a$ is the crack length, $\sigma_{ij}$, $\varepsilon_{ij}$ and $u_i$ are stress, strain and displacement components, respectively. $(\dot{\ })$ represents the temporal derivative $\left.\frac{\partial(\ )}{\partial t}\right|_{x_1, x_2}$. The term $\int_\Gamma n_j \sigma_{ij} \dot{u}_i d\Gamma$ is the power of the external force, and $\int_A \sigma_{ij} \dot{\varepsilon}_{ij} dA$ is the power of the internal force.

The power balance relation, Eq. (1), is simple, clear and correct. However, it is very interesting to find that for steady-state crack propagation in linear elastic materials, the surface-forming ERR $G_s$ cannot degenerate to the $J$-integral, as demonstrated as follows.

The stress field and the displacement field near a crack tip in plane stress condition are (Anderson, 2005)

$$\begin{Bmatrix} \sigma_{11} \\ \sigma_{12} \\ \sigma_{22} \end{Bmatrix} = \frac{K_I}{\sqrt{2\pi r}} \cos\frac{\theta}{2} \begin{Bmatrix} 1 - \sin\frac{\theta}{2}\sin\frac{3\theta}{2} \\ \sin\frac{\theta}{2}\cos\frac{3\theta}{2} \\ 1 + \sin\frac{\theta}{2}\sin\frac{3\theta}{2} \end{Bmatrix} \tag{2}$$

and

$$\begin{Bmatrix} u_1 \\ u_2 \end{Bmatrix} = \frac{K_I}{2\mu}\sqrt{\frac{r}{2\pi}} \begin{Bmatrix} \cos\frac{\theta}{2}\left(\kappa - 1 + 2\sin^2\frac{\theta}{2}\right) \\ \sin\frac{\theta}{2}\left(\kappa + 1 - 2\cos^2\frac{\theta}{2}\right) \end{Bmatrix} \tag{3}$$

where $r$ and $\theta$ are polar coordinates as shown in Fig. 1. $\mu = \frac{E}{2(1+\nu)}$ is the shear modulus and $\kappa = (3-\nu)/(1+\nu)$. $E$ and $\nu$ are Young's modulus and Poisson ratio, respectively. $K_I$ is the stress intensity factor for Mode I. The corresponding strain field can be calculated by the following constitutive equations,

$$\begin{Bmatrix} \varepsilon_{11} \\ \varepsilon_{12} \\ \varepsilon_{22} \end{Bmatrix} = \frac{1}{E} \begin{Bmatrix} \sigma_{11} - \nu\sigma_{22} \\ (1+\nu)\sigma_{12} \\ \sigma_{22} - \nu\sigma_{11} \end{Bmatrix} \tag{4}$$

For an arbitrary physical quantity $\Phi$, the following relation holds when a crack propagates in a steady-state way

$$\left.\frac{\partial \Phi}{\partial a}\right|_{x_1,x_2} = -\left.\frac{\partial \Phi}{\partial x_1}\right|_{x_2,a} \tag{5}$$

The surface-forming energy release rate in Eq. (1) can then be further derived as

$$\begin{aligned} G_s &= \int_\Gamma n_j \sigma_{ij} \frac{\partial u_i}{\partial a} d\Gamma - \int_A \sigma_{ij} \frac{\partial \varepsilon_{ij}}{\partial a} dA \\ &= -\int_\Gamma n_j \sigma_{ij} \frac{\partial u_i}{\partial x_1} d\Gamma + \int_A \sigma_{ij} \frac{\partial \varepsilon_{ij}}{\partial x_1} dA \end{aligned} \tag{6}$$

Substituting the stress field, the displacement field and the strain field of a crack tip of linear elastic materials into Eq. (6) yields

$$G_s = \frac{K_I^2}{4E}(3+\nu) \tag{7}$$

which is interestingly found to be smaller than $J$-integral $J = \frac{K_I^2}{E}$.

This inconsistence between the surface-forming ERR and $J$-integral for linear elastic fracture problem stimulates us to investigate the power balance in the second way and study the area within a comoving contour $\Gamma$ surrounding the crack tip as shown in Fig. 2. $x_1', x_2'$ is a comoving coordinate system (moving with the crack tip), and $A_{mov}$ is enclosed by the moving contour. At the initial moment $t_n$, two coordinate systems ($x_1', x_2'$ and $x_1, x_2$) coincide and the crack length is denoted by $a_n$. At a later moment $t$, we have the relationship

$$x_1' = x_1 - [a(t) - a_n], \quad x_2' = x_2 \tag{8}$$

where $a(t)$ is the corresponding crack length.

The corresponding power balance relation during crack propagation can be written as

$$G_L \dot{a} = \int_\Gamma n_j \sigma_{ij} \dot{u}_i d\Gamma + \dot{a} \int_\Gamma \hat{w}_0 n_1 d\Gamma - \frac{d}{dt} \int_{A_{mov}} \hat{w}_0 dA \tag{9}$$

where $G_L$ is named as the local energy release rate, $\hat{w}_0(t) = \int_0^t \sigma_{ij} \dot{\varepsilon}_{ij} dt$ is named as the accumulated work density, which is the same as the stress working density used by Brust and Atluri (1986). $t$ represents the time moment and moment 0 is the reference moment, at which material points have no stress and have not experienced any plastic deformation. The first term of RHS $\int_\Gamma n_j \sigma_{ij} \dot{u}_i d\Gamma$ is the power of the external force, and the second term $\dot{a}\int_\Gamma \hat{w}_0 n_1 d\Gamma$ is the power change due to the influx and outflow of material across the comoving contour. The third term $\frac{d}{dt}\int_{A_{mov}} \hat{w}_0 dA$ and $G_L \dot{a}$ represent the power inputs to the area and the crack, respectively. It should be pointed out that if there is no crack and singularity, the RHS of Eq. (9) can be proved to be zero as presented in Appendix A.

For a steady-state crack propagation in linear elastic materials, the accumulated work density $\hat{w}_0$ degenerates to the strain energy density $w$, and the local ERR

$$G_L = \int_\Gamma n_j \sigma_{ij} \frac{\partial u_i}{\partial a}\bigg|_{x_1,x_2} d\Gamma + \int_\Gamma w n_1 d\Gamma - \frac{d}{da}\int_{A_{mov}} \hat{w}_0 dA$$
$$= -\int_\Gamma n_j \sigma_{ij} \frac{\partial u_i}{\partial x_1} d\Gamma + \int_\Gamma w n_1 d\Gamma - \frac{d}{da}\int_{A_{mov}} \hat{w}_0 dA \qquad (10)$$
$$= -\int_\Gamma n_j \sigma_{ij} \frac{\partial u_i}{\partial x_1} d\Gamma + \int_\Gamma w n_1 d\Gamma$$

which is just the *J*-integral. $\frac{d}{dt}\int_{A_{mov}} \hat{w}_0 dA = 0$ is used in the above derivation due to the condition of steady-state.

Therefore, in some situations the local ERR $G_L$ and the surface-forming ERR $G_s$ can be different, although both of them are introduced by the power balance of the area within a contour. In the following, we first attempt to understand this interesting contradiction in a mathematical way. From Eq. (1), the surface-forming ERR can be further written as

$$G_s \dot{a} = \int_\Gamma n_j \sigma_{ij} \dot{u}_i d\Gamma - \int_A \sigma_{ij} \dot{\varepsilon}_{ij} dA$$
$$= \int_\Gamma n_j \sigma_{ij} \dot{u}_i d\Gamma - \int_A \frac{\partial \hat{w}_0}{\partial t}\bigg|_{x_1,x_2} dA$$
$$= \int_\Gamma n_j \sigma_{ij} \dot{u}_i d\Gamma + \int_A \dot{a}\frac{\partial \hat{w}_0}{\partial x_1} dA - \int_A \frac{\partial \hat{w}_0}{\partial t}\bigg|_{x'_1,x_2} dA \qquad (11)$$
$$= \int_\Gamma n_j \sigma_{ij} \dot{u}_i d\Gamma + \dot{a}\int_A \frac{\partial \hat{w}_0}{\partial x_1} dA - \frac{d}{dt}\int_{A_{mov}} \hat{w}_0 dA$$

The relations $\frac{\partial \hat{w}_0}{\partial t}\bigg|_{x_1,x_2} = \frac{\partial \hat{w}_0}{\partial t}\bigg|_{x'_1,x_2} - \dot{a}\frac{\partial \hat{w}_0}{\partial x_1}\bigg|_{x_2,t}$ and $\frac{\partial \hat{w}_0}{\partial t}\bigg|_{x_1,x_2} = \sigma_{ij}\dot{\varepsilon}_{ij}$ are used in the above derivation. Compare Eq. (9) with Eq. (11), we can see that only the second

terms of the RHS, $\dot{a}\int_A \frac{\partial \hat{w}_0}{\partial x_1}dA$ and $\dot{a}\int_\Gamma \hat{w}_0 n_1 d\Gamma$, seem different. If the Gauss theorem is applicable, $\int_A \frac{\partial \hat{w}_0}{\partial x_1}dA = \int_\Gamma \hat{w}_0 n_1 d\Gamma$, and $G_s = G_L$. However, since there is a singular point at the crack tip within the contour for a linear elastic fracture problem, the Gauss theorem fails, then $\int_A \frac{\partial \hat{w}_0}{\partial x_1}dA \ne \int_\Gamma \hat{w}_0 n_1 d\Gamma$ and $G_s \ne G_L$.

Based on the derivation above, it should be emphasized that Eq. (9) is not always equivalent to Eq. (1), but their equivalence is incorrectly adopted in our previous paper by Xiao et al. (2015) and also some fracture mechanics textbooks and lecture notes (Bower, 2005; Suo, 2016) on the derivation of *J*-integral by using the Reynolds transport theorem. The Reynolds transport theorem

$$\frac{d}{dt}\int_{A_{sta}} \hat{w}_0 dA = \frac{d}{dt}\int_{A_{mov}} \hat{w}_0 dA - \dot{a}\int_\Gamma \hat{w}_0 n_1 d\Gamma \tag{12}$$

involves the conversion between the integrations over a contour and an area, and therefore essentially uses the Gauss theorem. If there are singular points within the integration area, the applicability of the Reynolds transport theorem expressed by Eq. (12) and the Gauss theorem should be carefully checked. The correct transport relation for the contour with crack singularity should be

$$\frac{d}{dt}\int_{A_{sta}} \hat{w}_0 dA = \frac{d}{dt}\int_{A_{mov}} \hat{w}_0 dA - \dot{a}\int_\Gamma \hat{w}_0 n_1 d\Gamma + \dot{a}\int_{\Gamma_{tip}} \hat{w}_0 n_1 d\Gamma \tag{13}$$

The derivation can be found in Appendix B (see Eq. (B5)), and $\Gamma_{tip}$ is an infinitesimal contour surrounding the crack tip as shown in Fig. 3.

We have illustrated in a mathematical way that in some situations the local ERR $G_L$ and the surface-forming ERR $G_s$ may be different. In the next section, we will further discuss their physical meanings.

## 3 The formulae and physical meanings of the surface-forming ERR and the local ERR

In order to better illustrate the physical meanings of the two ERRs, their concise formulae are derived first.

### 3.1 The formulae of the surface-forming energy release rate $G_s$

In our previous paper (Xiao et al., 2015), the formulae of $G_s$ have been derived. Although Formula III and IV in that paper are correct, we also failed to distinguish the inequivalence between the ERRs from Eq. (1) and Eq. (9). In this subsection, we will derive Formula III and IV (called Formula S2 and S1 in this paper) directly without using Eq. (12).

According to Eq. (1), $G_s$ can be rewritten as

$$G_s = \int_\Gamma n_j \sigma_{ij} \frac{\dot{u}_i}{\dot{a}} d\Gamma - \int_A \sigma_{ij} \frac{\dot{\varepsilon}_{ij}}{\dot{a}} dA$$
$$= \int_\Gamma n_j \sigma_{ij} \frac{\partial u_i}{\partial a}\bigg|_{x_1,x_2} d\Gamma - \int_A \sigma_{ij} \frac{\partial \varepsilon_{ij}}{\partial a}\bigg|_{x_1,x_2} dA \tag{14}$$

The surface-forming energy release rate $G_s$ defined in Eq. (1) or Eq. (14) is a physical quantity with a clear meaning and should be independent of the contour. Therefore we may select an infinitesimal contour $\Gamma_{tip}$ as shown in Fig. 3(a), and its surrounded area is $A_{tip}$. Equation (14) can be rewritten as

$$G_s = \int_{\Gamma_{tip}} n_j \sigma_{ij} \frac{\partial u_i}{\partial a}\bigg|_{x_1,x_2} d\Gamma - \int_{A_{tip}} \sigma_{ij} \frac{\partial \varepsilon_{ij}}{\partial a}\bigg|_{x_1,x_2} dA$$
$$= \int_{\Gamma_{tip}} n_j \sigma_{ij} \frac{\partial u_i}{\partial a}\bigg|_{x_1',x_2} d\Gamma - \int_{\Gamma_{tip}} n_j \sigma_{ij} \frac{\partial u_i}{\partial x_1}\bigg|_{x_2,a} d\Gamma - \int_{A_{tip}} \sigma_{ij} \frac{\partial \varepsilon_{ij}}{\partial a}\bigg|_{x_1',x_2} dA + \int_{A_{tip}} \sigma_{ij} \frac{\partial \varepsilon_{ij}}{\partial x_1}\bigg|_{x_2,a} dA \tag{15}$$

In the following, we will illustrate that the first, the third and the fourth terms of the RHS approach to zero for an infinitesimal contour.

If $\sigma_{ij}\varepsilon_{ij}$ near the crack tip is on the order of $r^{-1}$ as in the K field or HRR field, $\sigma_{ij}\frac{\partial \varepsilon_{ij}}{\partial a}\bigg|_{x_1',x_2}$ is on the same order $r^{-1}$ or less singular, and $\sigma_{ij}\frac{\partial u_i}{\partial a}\bigg|_{x_1',x_2}$ is on the order of $r^0$. Obviously the first and the third terms of the RHS of Eq. (15)

$$\lim_{\Gamma_{tip}\to 0}\int_{\Gamma_{tip}} n_j \sigma_{ij} \frac{\partial u_i}{\partial a}\bigg|_{x_1',x_2} d\Gamma = 0 \tag{16}$$

$$\lim_{A_{tip}\to 0}\int_{A_{tip}} \sigma_{ij} \frac{\partial \varepsilon_{ij}}{\partial a}\bigg|_{x_1',x_2} dA = 0 \tag{17}$$

$\sigma_{ij}\frac{\partial \varepsilon_{ij}}{\partial x_1}\bigg|_{x_2,a}$ in the fourth term is on the order $r^{-2}$ or less singular, and is assumed to be expressed as $f(\theta)r^{-2}$ without losing generality. The fourth term of the RHS of Eq. (15) becomes

$$\lim_{A_{tip}\to 0}\int_{A_{tip}} \sigma_{ij} \frac{\partial \varepsilon_{ij}}{\partial x_1}\bigg|_{x_2,a} dA = \int_{-\pi}^{\pi} f(\theta) d\theta \lim_{R\to 0}\int_0^R \frac{1}{r^2} r dr \tag{18}$$

Noting that $\lim_{R \to 0} \int_0^R \frac{1}{r^2} r dr \to \infty$, if $\int_{-\pi}^{\pi} f(\theta) d\theta \neq 0$, $\lim_{A_{tip} \to 0} \int_{A_{tip}} \sigma_{ij} \frac{\partial \varepsilon_{ij}}{\partial x_1} \bigg|_{x_2, a} dA$ and the surface-forming ERR $G_s$ will become infinite according to Eqs. (15) and (18).

Based on Eq. (1), $G_s$ has a clear physical meaning and should not be infinite, therefore

$$\int_{-\pi}^{\pi} f(\theta) d\theta = 0 \tag{19}$$

This can also be validated by the crack tip field of linear elastic materials.

Substituting Eqs. (16)-(19) into Eq. (15) yields

**Formula S1**:

$$G_s = -\int_{\Gamma_{tip}} n_j \sigma_{ij} \frac{\partial u_i}{\partial x_1} d\Gamma \tag{20}$$

which is the same as Formula IV in our previous paper, and can be derived in another way for hyperelastic materials as shown in Appendix B.

Formula S1 is applicable for an infinitesimal contour. For a finite contour $\Gamma$ as shown in Fig. 3(b), another two auxiliary contours tightly around the crack surfaces $\Gamma_+$ and $\Gamma_-$ can be introduced to form an enclosed contour $(\Gamma + \Gamma_+ + \Gamma_{tip}^- + \Gamma_-)$ without a singular point in it. The Gauss theorem is then used to obtain the following relation

$$\int_{A-A_{tip}} \sigma_{ij} \frac{\partial \varepsilon_{ij}}{\partial x_1} dA$$
$$= \int_{\Gamma + \Gamma_+ + \Gamma_{tip}^- + \Gamma_-} n_j \sigma_{ij} \frac{\partial u_i}{\partial x_1} d\Gamma \tag{21}$$
$$= \int_{\Gamma} n_j \sigma_{ij} \frac{\partial u_i}{\partial x_1} d\Gamma + \int_{\Gamma_+} n_j \sigma_{ij} \frac{\partial u_i}{\partial x_1} d\Gamma + \int_{\Gamma_{tip}^-} n_j \sigma_{ij} \frac{\partial u_i}{\partial x_1} d\Gamma + \int_{\Gamma_-} n_j \sigma_{ij} \frac{\partial u_i}{\partial x_1} d\Gamma$$

It is assumed that no external force is applied on the crack surfaces within the contour, then $n_j \sigma_{ij} = 0$ on the contour $\Gamma_+$ and $\Gamma_-$, so

$$\int_{\Gamma_+} n_j \sigma_{ij} \frac{\partial u_i}{\partial x_1} d\Gamma + \int_{\Gamma_-} n_j \sigma_{ij} \frac{\partial u_i}{\partial x_1} d\Gamma = 0 \tag{22}$$

Substituting the above equation into Eq. (21) and noting Eq. (20), we obtain the formula of the surface-forming ERR $G_s$ for a finite contour

**Formula S2:**
$$G_s = -\int_{\Gamma} n_j \sigma_{ij} \frac{\partial u_i}{\partial x_1} d\Gamma + \int_{A-A_{tip}} \sigma_{ij} \frac{\partial \varepsilon_{ij}}{\partial x_1} dA \tag{23}$$

which is the same as the Formula III in our previous paper.

The formulae of $G_s$ are concise and only expressed by the deformation and stress status at the current moment, and therefore are easy to use.

## 3.2 The formulae of the local energy release rate $G_L$

According to Eq. (9), $G_L$ for a finite contour can be rewritten as

$$G_L = \int_\Gamma n_j \sigma_{ij} \frac{\partial u_i}{\partial a}\bigg|_{x_1,x_2} d\Gamma + \int_\Gamma \hat{w}_0 n_1 d\Gamma - \frac{d}{da}\int_{A_{mov}} \hat{w}_0 dA$$

$$= \int_\Gamma \left(\hat{w}_0 n_1 - n_j \sigma_{ij} \frac{\partial u_i}{\partial x_1}\bigg|_{x_2,a}\right) d\Gamma + \left[\int_\Gamma n_j \sigma_{ij} \frac{\partial u_i}{\partial a}\bigg|_{x_1',x_2} d\Gamma - \frac{d}{da}\int_{A_{mov}} \hat{w}_0 dA\right] \quad (24)$$

With the help of an enclosed contour $\Gamma + \Gamma_+ + \Gamma_{tip}^- + \Gamma_-$ as shown in Fig. 3(b), the terms in the second bracket of Eq. (24) can be further simplified as follows

$$\int_\Gamma n_j \sigma_{ij} \frac{\partial u_i}{\partial a}\bigg|_{x_1',x_2} d\Gamma - \frac{d}{da}\int_{A_{mov}} \hat{w}_0 dA$$

$$= \int_{\Gamma+\Gamma_+ +\Gamma_{tip}^- +\Gamma_-} n_j \sigma_{ij} \frac{\partial u_i}{\partial a}\bigg|_{x_1',x_2} d\Gamma - \int_{\Gamma_+} n_j \sigma_{ij} \frac{\partial u_i}{\partial a}\bigg|_{x_1',x_2} d\Gamma - \int_{\Gamma_-} n_j \sigma_{ij} \frac{\partial u_i}{\partial a}\bigg|_{x_1',x_2} d\Gamma \quad (25)$$

$$+ \int_{\Gamma_{tip}} n_j \sigma_{ij} \frac{\partial u_i}{\partial a}\bigg|_{x_1',x_2} d\Gamma - \frac{d}{da}\int_{A_{mov}-A_{tip}^{mov}} \hat{w}_0 dA - \frac{d}{da}\int_{A_{tip}^{mov}} \hat{w}_0 dA$$

The crack surfaces ($\Gamma_+$ and $\Gamma_-$) are assumed to be traction free such that

$$\int_{\Gamma_+} n_j \sigma_{ij} \frac{\partial u_i}{\partial a}\bigg|_{x_1',x_2} d\Gamma + \int_{\Gamma_-} n_j \sigma_{ij} \frac{\partial u_i}{\partial a}\bigg|_{x_1',x_2} d\Gamma = 0 \quad (26)$$

$\int_{A_{tip}^{mov}} \hat{w}_0 dA$ should also approach to zero for an infinitesimal contour $\Gamma_{tip}$, otherwise there will be infinite average accumulated work density which is physically unreasonable. For example, the singularity of the strain energy density (i.e. a special case of $\hat{w}_0$) for HRR field and K field is on the order of $r^{-1}$ such that $\lim_{A_{tip}^{mov} \to 0} \int_{A_{tip}^{mov}} \hat{w}_0 dA \to 0$. Noting that the area $A_{tip}^{mov}$ and the crack extension $\Delta a$ are two independent infinitesimal quantities, we then obtain

$$\frac{d}{da}\int_{A_{tip}^{mov}} \hat{w}_0 dA = 0 \quad (27)$$

Substituting Eqs. (16) (26) and (27) into Eq. (25) yields

$$\int_{\Gamma} n_j \sigma_{ij} \left.\frac{\partial u_i}{\partial a}\right|_{x_1', x_2} d\Gamma - \frac{d}{da}\int_{A_{mov}} \hat{w}_0 dA$$

$$= \int_{\Gamma+\Gamma_++\Gamma_{tip}^-+\Gamma_-} n_j \sigma_{ij} \left.\frac{\partial u_i}{\partial a}\right|_{x_1', x_2} d\Gamma - \frac{d}{da}\int_{A_{mov}-A_{tip}^{mov}} \hat{w}_0 dA$$

$$= \int_{A-A_{tip}} \sigma_{ij} \left.\frac{\partial \varepsilon_{ij}}{\partial a}\right|_{x_1', x_2} dA - \int_{A-A_{tip}} \left.\frac{\partial \hat{w}_0}{\partial a}\right|_{x_1', x_2} dA \qquad (28)$$

$$= \int_{A-A_{tip}} \left[\left(\sigma_{ij}\left.\frac{\partial \varepsilon_{ij}}{\partial a}\right|_{x_1,x_2} - \left.\frac{\partial \hat{w}_0}{\partial a}\right|_{x_1,x_2}\right) + \left(\sigma_{ij}\left.\frac{\partial \varepsilon_{ij}}{\partial x_1}\right|_{x_2,a} - \left.\frac{\partial \hat{w}_0}{\partial x_1}\right|_{x_2,a}\right)\right] dA$$

$$= \int_{A-A_{tip}} \left(\sigma_{ij}\left.\frac{\partial \varepsilon_{ij}}{\partial x_1}\right|_{x_2,a} - \left.\frac{\partial \hat{w}_0}{\partial x_1}\right|_{x_2,a}\right) dA$$

The relations $\sigma_{ij}\left.\frac{\partial \varepsilon_{ij}}{\partial a}\right|_{x_1,x_2} - \left.\frac{\partial \hat{w}_0}{\partial a}\right|_{x_1,x_2} = \frac{1}{\dot{a}}\left(\sigma_{ij}\left.\frac{\partial \varepsilon_{ij}}{\partial t}\right|_{x_1,x_2} - \left.\frac{\partial \hat{w}_0}{\partial t}\right|_{x_1,x_2}\right) = 0$,

$\left.\frac{\partial(\cdot)}{\partial a}\right|_{x_1',x_2} = \left.\frac{\partial(\cdot)}{\partial a}\right|_{x_1,x_2} + \left.\frac{\partial(\cdot)}{\partial x_1}\right|_{x_2,a}$, and the Gauss theorem are used in the above derivation.

Substituting Eq. (28) into Eq. (24) yields the formula of the local ERR $G_L$ for a finite contour

**Formula L1:**

$$\begin{aligned} G_L &= \int_{\Gamma}\left(\hat{w}_0 n_1 - n_j \sigma_{ij}\frac{\partial u_i}{\partial x_1}\right) d\Gamma + \int_{A-A_{tip}}\left(\sigma_{ij}\frac{\partial \varepsilon_{ij}}{\partial x_1} - \frac{\partial \hat{w}_0}{\partial x_1}\right) dA \\ &= \int_{\Gamma}\left(n_1 \int_0^t \sigma_{ij}\dot{\varepsilon}_{ij} dt - n_j \sigma_{ij}\frac{\partial u_i}{\partial x_1}\right) d\Gamma + \int_{A-A_{tip}}\left[\sigma_{ij}\frac{\partial \varepsilon_{ij}}{\partial x_1} - \frac{\partial\left(\int_0^t \sigma_{ij}\dot{\varepsilon}_{ij} dt\right)}{\partial x_1}\right] dA \end{aligned} \qquad (29)$$

which is the same as the path-independent integral proposed by Brust et al. (1986).

For an infinitesimal contour $\Gamma_{tip}$, the second integral vanishes and we have another formula

**Formula L2:** $\quad G_L = \int_{\Gamma_{tip}}\left(\hat{w}_0 n_1 - n_j \sigma_{ij}\frac{\partial u_i}{\partial x_1}\right) d\Gamma = \int_{\Gamma_{tip}}\left(n_1 \int_0^t \sigma_{ij}\dot{\varepsilon}_{ij} dt - n_j \sigma_{ij}\frac{\partial u_i}{\partial x_1}\right) d\Gamma \qquad (30)$

Both formulae of $G_L$ actually include double integrals, in which the whole deformation history of material points is needed to calculate the accumulated work density $\hat{w}_0 = \int_0^t \sigma_{ij}\dot{\varepsilon}_{ij} dt$, which makes the calculation hard to implement. It is noted from Formula L2 that $G_L$ can degenerate into the *J*-integral for a crack in hyperelastic materials, but $G_L$ is also applicable to the crack propagation in elastic-plastic materials.

### 3.3 The physical meanings of $G_s$ and $G_L$

To exhibit the physical meanings of the surface-forming ERR $G_s$ and the local ERR $G_L$, we first study a crack propagation with a cohesive zone at the crack tip as shown in Fig. 4. The contour $\Gamma_{tip}$ is chosen as the contour tightly around the cohesive zone, i.e. $\Gamma_{cz}$, and we adopt Formula S1 and Formula L2 to calculate $G_s$ and $G_L$ respectively.

$$G_s = -\int_{\Gamma_{cz}} n_j \sigma_{ij} \frac{\partial u_i}{\partial x_1} d\Gamma = -\int_{x_{tip}}^{x_{tip}+l_{cz}} \sigma(x_1) \frac{\partial \delta(x_1)}{\partial x_1} dx_1 = \int_0^{\delta_{tip}} \sigma(\delta) d\delta \tag{31}$$

$$G_L = \int_{\Gamma_{cz}} \left( \hat{w}_0 n_1 - n_j \sigma_{ij} \frac{\partial u_i}{\partial x_1} \right) d\Gamma = -\int_{\Gamma_{cz}} n_j \sigma_{ij} \frac{\partial u_i}{\partial x_1} d\Gamma = \int_0^{\delta_{tip}} \sigma(\delta) d\delta \tag{32}$$

where $x_{tip}$ is the coordinate of the crack tip, $l_{cz}$ is the length of the cohesive zone, $\delta$ and $\sigma$ are the opening displacement and traction in the cohesive zone, $\delta_{tip}$ is the crack opening displacement at the crack tip as shown in Fig. 4.

It is very interesting to note that $G_s = G_L$ for a crack propagation with a cohesive zone, but $G_s < G_L$ for linear elastic fracture problems as we have demonstrated in Section 2. We attribute this inconsistence to the existence of the stress singularity at a crack tip and will disclose its effect on the energy or power.

Figure 5 shows a propagating crack with a stress singularity at the tip. Although the crack grows very slowly, the hoop stress of a material point closely ahead of the crack tip (marked as a red dot) will drop suddenly and dramatically from almost infinite to zero when the crack passes through it, which will result in a stress wave and a nonzero kinetic energy. This phenomenon can be observed sometimes. For brittle materials under quasi-static loading, this wave disturbance usually leads to discontinuous crack propagation, while a crack propagates smoothly in plastic materials due to smaller stress drop and smaller wave disturbance at the crack tip. Therefore, the accompanying kinetic energy with crack propagation is closely related to the stress singularity at the tip.

In the most general situation, the energy dissipation during the crack propagation in elastic-plastic materials should include three parts: the energy to form the new surfaces $U_S$, the plastic dissipation energy $U_P$ and the accompanying kinetic energy $U_K$ at the crack tip. $U_P$ is loading-mode-dependent as discussed in our previous paper (Xiao et al., 2015), and has been excluded in $G_s$ and $G_L$. Moreover, our investigation above indicates that if there is a stress singularity and the resulting accompanying kinetic energy $U_K$, $G_s < G_L$; Otherwise, there is no accompanying kinetic energy and $G_s = G_L$. We can then reach the following conclusions:

1) The surface-forming energy release rate $G_s$ represents the driving force to overcome $U_S$;

2) The local energy release rate $G_L$ represents the driving force to overcome $U_S + U_K$;

3) The traditional (or global) energy release rate $G$ represents the driving force to overcome $U_S + U_K + U_P$.

Here for completeness, we also provide the following formula for the traditional/global ERR $G$

$$G = \int_\Gamma n_j \sigma_{ij} \left.\frac{\partial u_i}{\partial a}\right|_{x_1,x_2} d\Gamma + \int_\Gamma w_e n_1 d\Gamma - \frac{d}{da}\int_{A_{mov}} w_e dA \qquad (33)$$

where $w_e$ is the elastic strain energy density as schematically shown in Fig. 6. It should be pointed out that to correctly obtain the global ERR $G$, the contour $\Gamma$ must enclose the plastic loading zone (see Fig. 7); Otherwise, the integral will be path-dependent. For a steady-state crack propagation in elastic-plastic materials, such as the case shown in Fig. 8, many researchers have obtained the global ERR $G$ by computing the difference between the elastic strain energy of the region far ahead the crack tip and that of far behind it, which is essentially the special case of Eq. (33) by adopting the contour $\Gamma_1 + \Gamma_2 + \Gamma_3 + \Gamma_4 + \Gamma_5$ shown in Fig. 8

Regarding the inconsistence between the local ERR $G_L$ and the energy dissipation related to the surface forming, another interpretation is given as follows. Figure 9(a) shows a crack in a linear elastic material, and K-field of mode I type exists near the crack tip. In this case, $G_L$ degenerates to the $J$-integral. In many literature, such as the textbook by Anderson (2005), $J$-integral (or $G_L$) can be obtained by investigating a virtual crack growth process. The initial length of the crack is $a$. One can imagine releasing the traction $F_y(x)$ of the region between $x=0$ to $x=\Delta a$ and the corresponding work required to advance the crack by $\Delta a$ is

$$\Delta U = \int_0^{\Delta a} 2 \times \frac{1}{2} F_y(x) u_y(x-\Delta a) dx$$
$$= \int_0^{\Delta a} \sigma_{yy}(x) u_y(x-\Delta a) dx \tag{34}$$

$\sigma_{yy}(x)$ and $2u_y(x-\Delta a)$ are the stress and opening displacement shown in Figure 9(a). The factor of $\frac{1}{2}$ means the linear relationship between the traction $F_y(x)$ and $u_y(x-\Delta a)$. For a point at $x$, the work can be represented by the area of the blue shadow in Fig. 9(c). $J$-integral (or $G_L$) is then obtained by

$$J = G_L = \lim_{\Delta a \to 0}\left(\frac{\Delta U}{\Delta a}\right) = \lim_{\Delta a \to 0} \frac{1}{\Delta a} \int_0^{\Delta a} \sigma_{yy}(x) u_y(x-\Delta a) dx \tag{35}$$

If this treatment is applied to a crack with a cohesive zone in linear elastic materials, it will be found that the work is not consistent with the corresponding work from a cohesive model as shown in Fig. 9(b). We investigate a material point on the crack plane (denoted by the red point) during the crack propagation, and Fig. 9(c) exhibits its force-displacement curve when the crack propagates. The area of the shadow under the red curve is actually the work required to form the new surfaces $U_S$, i.e., the critical energy defined in the cohesive law, and is equal to the surface-forming energy release rate $G_s$. Obviously, the area under the red cohesive curve $U_S$ is not necessarily equal to the area under the blue line $G_L$. Considering that Eq. (34) is originally used in linear elastic materials without a cohesive zone, we shorten the cohesive zone to the crack tip by increasing the cohesive strength to infinite. We believe that the local ERR $G_L$ is still different from $U_S$.

Various energy-related fracture parameters or criteria discussed in this paper are summarized in Table 1

**Table 1** Summary of various energy-related fracture parameters or criteria for crack propagation in elastic-plastic materials.

| Fracture parameters or criteria | Applic-ability to elastic-plastic crack propagation | Physical meaning | Advantages | Disadvantages |
|---|---|---|---|---|
| The surface-forming ERR $$G_s = -\int_{\Gamma_{tip}} n_j \sigma_{ij} \frac{\partial u_i}{\partial x_1} d\Gamma$$ | Yes | Driving force to overcome the surface energy $U_S$ | Easy to compute | Cannot degenerate to the $J$-integral if stress singularity exists |
| The local ERR $$G_L = \int_{\Gamma_{tip}} \left( n_1 \int_0^t \sigma_{ij} \dot{\varepsilon}_{ij} dt - n_j \sigma_{ij} \frac{\partial u_i}{\partial x_1} \right) d\Gamma$$ | Yes | Driving force to overcome the surface energy $U_S$ and the kinetic energy $U_K$ | Can degenerate to the $J$-integral | Double integral involving integration over deformation history, hard to calculate |
| The global ERR $$G = \int_{\Gamma} n_j \sigma_{ij} \frac{\partial u_i}{\partial a}\bigg|_{x_1,x_2} d\Gamma + \int_{\Gamma} w_e n_1 d\Gamma - \frac{d}{da}\int_{A_{mov}} w_e dA$$ | Yes | Driving force to overcome the surface energy $U_S$, the kinetic energy $U_K$ and the plastic dissipation $U_P$ | | $G_c$ is loading-mode-dependent since it includes plastic dissipation |
| The $J$-integral $$J = \int_{\Gamma} \left( wn_1 - n_j \sigma_{ij} \frac{\partial u_i}{\partial x_1} \right) d\Gamma$$ | No | N/A | | |

# 4 Discussions on the applications of different fracture criteria for elastic-plastic crack propagation.

Several issues on determining and simulating crack propagation in elastic-plastic materials need our attention.

For elastic-perfectly plastic materials, the strain at a propagating crack tip has a weaker singularity on the order of $\ln(r)$ and the stress has no singularity (Rice 1966), therefore the J-integral, the surface-forming ERR, and the local ERR on an infinitesimal contour become zero, as pointed out in Rice paradox (Kfouri and Miller 1976; Kfouri and Rice 1977; Rice 1978). In our previous paper, this paradox was interpreted by noting that a cohesive zone for a crack in elastic-perfectly plastic materials cannot shrink to a point and therefore an infinitesimal integration contour is invalid. But the problem has not been solved completely. In this section, we will discuss how to determine and simulate a propagating crack in elastic-perfectly plastic materials.

We first investigate the applicability of the cohesive zone model through simulations by finite element method software ABAQUS (ABAQUS, 2014). A test example is a strip with a horizontal straight crack located in the left region subjected to vertical tensile displacement loading as shown in Fig. 10(a). The cohesive elements with a cohesive strength $\sigma_c$ are deployed along the crack plane. The other regions are an elastic-perfectly plastic material with a yielding stress $\sigma_y$. The crack propagation process then can be simulated. If we select $\sigma_c > \sigma_y$, no matter how large the loading is, the crack will not propagate as shown in Fig. 10(b) since the stress level outside the cohesive zone cannot reach the cohesive strength; If $\sigma_c < \sigma_y$, there will be no plastic deformation outside the cohesive zone as in Fig. 10(c). Both simulation results are unreasonable. Therefore, the cohesive strength $\sigma_c$ must be exactly the same as the yielding stress $\sigma_y$, which is difficult to realize in simulations due to the numerical error. In other words, the simulated fracture behaviors are very sensitive to the parameters of the cohesive model, and therefore the cohesive model based fracture criterion is not a proper one for elastic-perfectly plastic materials.

We then switch to the surface-forming ERR based fracture criterion since it is easy to compute among all ERRs. According to the Rice Paradox, an infinitesimal contour should be avoided to use. An approximate ERR

$$\tilde{G}_s = -\int_{\Gamma_R} n_j \sigma_{ij} \frac{\partial u_i}{\partial x_1} d\Gamma \tag{36}$$

is then introduced on a finite circle contour $\Gamma_R$ as shown in Fig. 11, and the radius is $R$. Obviously, $\tilde{G}_s$ becomes $G_s$ when $R$ approaches zero. The relation between $\tilde{G}_s$ and $R$ for elastic-perfectly plastic materials is schematically demonstrated in Fig. 12(a). We will illustrate that when the contour is near the crack tip, the function curve will approach an asymptotic line passing through the origin, i.e. $\tilde{G}_s$ is in proportion to $R$. Noting that $\tilde{G}_s$ is the first term of Formula S2 and $G_s = 0$, we can use the second term $\int_{A-A_{tip}} \sigma_{ij} \frac{\partial \varepsilon_{ij}}{\partial x_1} dA$ to evaluate the order of $\tilde{G}_s$. It has been stated that $\sigma_{ij} \propto r^0$ and $\varepsilon_{ij} \propto \ln(r)$ for a crack propagation in elastic-perfectly materials (Rice 1966), then we can get $\sigma_{ij} \frac{\partial \varepsilon_{ij}}{\partial x_1} \propto \frac{1}{r}$ and

$$\int_{A-A_{tip}} \sigma_{ij} \frac{\partial \varepsilon_{ij}}{\partial x_1} dA = \int_{-\pi}^{\pi} \int_0^R \sigma_{ij} \frac{\partial \varepsilon_{ij}}{\partial x_1} r dr d\theta \propto R \tag{37}$$

Therefore $\tilde{G}_s \propto R$.

We may use the curve between $\tilde{G}_s$ and $R$ to determine whether a crack will propagate. A simple way is using the slope of the asymptotic line in Fig. 12(a), i.e. $g_s^d = \frac{d\tilde{G}_s}{dR}$. If it is larger than its critical value $g_{sc}^d$, the crack will propagate. This can be viewed as an extended surface-forming ERR based fracture criterion.

Although the surface-forming ERR $G_s$ for a propagating crack with a singular stress field at the tip (such as strain hardening materials) has a finite and reasonable value for an infinitesimal contour, its simulation or measurement error may be reduced if the relation between $\tilde{G}_s$ and $R$ is obtained first. In these materials, when a crack propagates, $\sigma_{ij}\varepsilon_{ij} \propto r^{-1}$ near the tip, and $\sigma_{ij}\dfrac{\partial \varepsilon_{ij}}{\partial x_1} \propto \dfrac{1}{r^2}$. Noting $\lim\limits_{A_{tip} \to 0}\int_{A_{tip}} \sigma_{ij}\dfrac{\partial \varepsilon_{ij}}{\partial x_1}dA = 0$ according to Eqs. (18) and (19), the relation curve between $\tilde{G}_s$ and $R$ should approach a horizontal asymptotic line when the contour near the tip as schematically shown in Fig. 12(b).

## 5 Conclusions

Two energy release rates applicable to a crack propagation in elastic-plastic materials are derived based on the power balance of the areas within a stationary contour and a comoving contour, respectively. Although both ERRs are reasonable, there are some interesting contradictions between them, which inspire this study on the underlying mechanisms. The following conclusions are reached:

(1) For a crack propagation with stress singularity at the tip, the surface-forming ERR $G_s$ is less than the local ERR $G_L$, and the latter can degenerate into the *J*-integral for a crack in hyperelastic materials. If there is no stress singularity, $G_s = G_L$.

(2) It can be deduced that the stress singularity leads to an accompanying kinetic energy at the crack tip. The local ERR $G_L$ represents the driving force to overcome the surface energy and the accompanying kinetic energy. The surface-forming ERR $G_s$ represents the driving force to overcome the surface energy only.

(3) For a crack propagation in elastic-plastic materials, we recommend using the surface-forming ERR $G_s$ based fracture criterion, since it has a wide applicability and concise formulae which are easy to compute among all ERRs. Especially in the case of elastic-perfectly plastic materials, even a cohesive model fails to predict reasonable fracture behaviors.

Acknowledgement

We would like to thank the Prof. Chad M Landis from the University of Texas at Austin for the discussion with him that inspired us to carry out this work. Besides, the support from National Natural Science Foundation of China (Grant Nos. 51232004, 11425208 and 11372158) and Tsinghua University Initiative Scientific Research Program (No. 2011Z02173) is also acknowledged.

**Appendix A**

For an enclosed contour with no crack and singularity in it,

$$\int_{\Gamma} n_j \sigma_{ij} \dot{u}_i d\Gamma + \dot{a} \int_{\Gamma} \hat{w}_0 n_1 d\Gamma - \frac{d}{dt} \int_{A_{mov}} \hat{w}_0 dA$$

$$= \int_{A} \sigma_{ij} \dot{u}_{i,j} d\Gamma + \dot{a} \int_{A} \frac{\partial \hat{w}_0}{\partial x_1} dA - \frac{d}{dt} \int_{A_{mov}} \hat{w}_0 dA$$

$$= \int_{A} \sigma_{ij} \dot{\varepsilon}_{ij} d\Gamma + \dot{a} \int_{A} \frac{\partial \hat{w}_0}{\partial x_1} dA - \int_{A} \left. \frac{\partial \hat{w}_0}{\partial t} \right|_{x_1', x_2} dA \qquad (A1)$$

$$= \int_{A} \left. \frac{\partial \hat{w}_0}{\partial t} \right|_{x_1, x_2} d\Gamma + \dot{a} \int_{A} \frac{\partial \hat{w}_0}{\partial x_1} dA - \int_{A} \left. \frac{\partial \hat{w}_0}{\partial t} \right|_{x_1', x_2} dA$$

$$= 0$$

**Appendix B**

For a continuous enclosed contour without any singular points in it, the following transport relation between a stationary coordinate system and a comoving coordinate system holds,

$$\frac{d}{dt}\int_{A_{sta}} \hat{w}_0 dA = \frac{d}{dt}\int_{A_{mov}} \hat{w}_0 dA - \dot{a}\int_{\Gamma} \hat{w}_0 n_1 d\Gamma \tag{B1}$$

If a crack tip singularity exists in the contour $\Gamma$ and the corresponding area A, we can introduce an infinitesimal contour $\Gamma_{tip}^-$ and two auxiliary contours tightly around the crack surfaces $\Gamma_+$ and $\Gamma_-$ to form an enclosed contour ($\Gamma + \Gamma_+ + \Gamma_{tip}^- + \Gamma_-$) without a singular point in it as showed in Fig. 3, and the corresponding area is $A - A_{tip}$. Then Eq. (B1) can be utilized as

$$\frac{d}{dt}\int_{A_{sta}-A_{sta}^{tip}} \hat{w}_0 dA = \frac{d}{dt}\int_{A_{mov}-A_{mov}^{tip}} \hat{w}_0 dA - \dot{a}\int_{\Gamma+\Gamma_++\Gamma_{tip}^-+\Gamma_-} \hat{w}_0 n_1 d\Gamma \tag{B2}$$

Noting that $n_1 = 0$ holds on $\Gamma_+$ and $\Gamma_-$, Eq. (B2) can be rewritten as

$$\begin{aligned}&\frac{d}{dt}\int_{A_{sta}} \hat{w}_0 dA - \frac{d}{dt}\int_{A_{sta}^{tip}} \hat{w}_0 dA \\ &= \frac{d}{dt}\int_{A_{mov}} \hat{w}_0 dA - \frac{d}{dt}\int_{A_{mov}^{tip}} \hat{w}_0 dA - \dot{a}\int_{\Gamma} \hat{w}_0 n_1 d\Gamma + \dot{a}\int_{\Gamma_{tip}} \hat{w}_0 n_1 d\Gamma\end{aligned} \tag{B3}$$

The accumulated work density $\hat{w}_0(t) = \int_0^t \sigma_{ij}\dot{\varepsilon}_{ij} dt \sim \sigma_{ij}\varepsilon_{ij}$ is on the order of $r^{-1}$ as in the K field or HRR field near the crack tip. The second term in the left hand side of Eq. (B3) $\frac{d}{dt}\int_{A_{sta}^{tip}} \hat{w}_0 dA = \int_{A_{sta}^{tip}} \left.\frac{\partial \hat{w}_0}{\partial t}\right|_{x_1,x_2} dA = \int_{A_{sta}^{tip}} \dot{a}\left.\frac{\partial \hat{w}_0}{\partial a}\right|_{x_1,x_2} dA$, and $\left.\frac{\partial \hat{w}_0}{\partial a}\right|_{x_1,x_2} \sim -\left.\frac{\partial \hat{w}_0}{\partial x_1}\right|_{x_2,a}$ is on the order of $r^{-2}$. Without losing generality, we can assume $\left.\frac{\partial \hat{w}_0}{\partial t}\right|_{x_1,x_2}$ to be expressed as $f(\theta)r^{-2}$ and

$$\lim_{A_{sta}^{tip}\to 0}\int_{A_{sta}^{tip}} \left.\frac{\partial \hat{w}_0}{\partial t}\right|_{x_1,x_2} dA = \int_{-\pi}^{\pi} f(\theta)d\theta \lim_{R\to 0}\int_0^R \frac{1}{r^2} rdr \tag{B4}$$

Noting that $\lim_{R\to 0}\int_0^R \frac{1}{r^2} rdr \to \infty$, if the coefficient term $\int_{-\pi}^{\pi} f(\theta)d\theta \neq 0$ and then $\lim_{A_{sta}^{tip}\to 0}\int_{A_{sta}^{tip}} \left.\frac{\partial \hat{w}_0}{\partial t}\right|_{x_1,x_2} dA$ will also tend to infinite, which is obviously unreasonable.

Therefore $\int_{-\pi}^{\pi} f(\theta)d\theta = 0$ and then $\frac{d}{dt}\int_{A_{sta}^{tip}} \hat{w}_0 dA = 0$.

For the second term in the right hand side of Eq. (B3), $\frac{d}{dt}\int_{A_{mov}^{tip}} \hat{w}_0 dA = \int_{A_{mov}^{tip}} \frac{\partial \hat{w}_0}{\partial t}\bigg|_{x_1',x_2} dA$ and $\frac{\partial \hat{w}_0}{\partial t}\bigg|_{x_1',x_2} \sim \frac{1}{r}$, therefore $\frac{d}{dt}\int_{A_{mov}^{tip}} \hat{w}_0 dA$ also tends to zero. Therefore, the correct transport relation for the contour including a crack tip singularity should be

$$\frac{d}{dt}\int_{A_{sta}} \hat{w}_0 dA = \frac{d}{dt}\int_{A_{mov}} \hat{w}_0 dA - \dot{a}\int_{\Gamma} \hat{w}_0 n_1 d\Gamma + \dot{a}\int_{\Gamma_{tip}} \hat{w}_0 n_1 d\Gamma \tag{B5}$$

Compare Eq. (B1) and Eq. (B5), we can see that their difference is the term $\dot{a}\int_{\Gamma_{tip}} \hat{w}_0 n_1 d\Gamma$, which is often omitted by many scholars in previous derivations of the *J*-integral, even in course lecture notes (Bower, 2005; Suo, 2016). It is easily found that usually $\int_{\Gamma_{tip}} \hat{w}_0 n_1 d\Gamma \neq 0$, which can be demonstrated by a linear elastic crack tip field.

For a crack in hyperelastic materials, the accumulated work density degenerates to the strain energy density, i.e. $\hat{w}_0 = w$. According to the correct transport relation Eq. (B5), the definition of $G_s$ Eq. (1) can be further written as

$$\begin{aligned} G_s \dot{a} &= \int_{\Gamma} n_j \sigma_{ij} \dot{u}_i d\Gamma - \int_A \sigma_{ij} \dot{\varepsilon}_{ij} dA \\ &= \int_{\Gamma} n_j \sigma_{ij} \dot{u}_i d\Gamma - \int_A \frac{\partial w}{\partial t}\bigg|_{x_1,x_2} dA \\ &= \dot{a}\left(\frac{1}{\dot{a}}\int_{\Gamma} n_j \sigma_{ij} \dot{u}_i d\Gamma + \int_{\Gamma} w n_1 d\Gamma - \frac{1}{\dot{a}}\frac{d}{dt}\int_{A_{mov}} w dA\right) - \dot{a}\int_{\Gamma_{tip}} w n_1 d\Gamma \end{aligned} \tag{B6}$$

The term in the bracket of the above equation is actually the *J*-integral, and the derivation can be found in Eqs. (24)-(30). Then the surface-forming ERR $G_s$ becomes

$$G_s = \int_{\Gamma_{tip}} \left(w n_1 - n_j \sigma_{ij} \frac{\partial u_i}{\partial x_1}\bigg|_{x_2,t}\right) d\Gamma - \int_{\Gamma_{tip}} w n_1 d\Gamma = -\int_{\Gamma_{tip}} n_j \sigma_{ij} \frac{\partial u_i}{\partial x_1} d\Gamma \tag{B7}$$

which is the same as Formula S1 in the following derivation of this paper.

**Figure captions**

Figure 1 Schematic of a stationary contour surrounding the crack tip.

Figure 2 Stationary and moving coordinate systems and corresponding contours

Figure 3 Schematic diagrams of (a) an infinitesimal contour closely surrounding the crack tip and (b) an enclosed integration contour.

Figure 4 Schematics of (a) a cohesive zone and (b) its cohesive law.

Figure 5 Schematics of a material point on the crack plane before and after the crack passing through it with dramatic stress drop.

Figure 6 Schematic of the elastic strain energy density.

Figure 7 Schematic an integration contour enclosing the plastic loading zone for the global energy release rate $G$

Figure 8 A cracked strip with constant displacement loading on upper and lower boundaries and an integration contour.

Figure 9 Schematic snapshots of a propagating crack (a) in a linear elastic material and (b) in a material with a cohesive zone; (c) The corresponding works required.

Figure 10 (a) Schematic of a strip subject to constant displacement loading on the upper and lower boundaries with initial crack and cohesive elements; (b)Simulation results of fracture for $\sigma_c > \sigma_y$ and (c) $\sigma_c < \sigma_y$.

Figure 11 Schematic of a finite circle contour

Figure 12 Schematics of the relation between $\tilde{G}_s$ and $R$ for crack propagation in (a) elastic-perfectly plastic materials and (b) materials with a singular stress field at the tip.

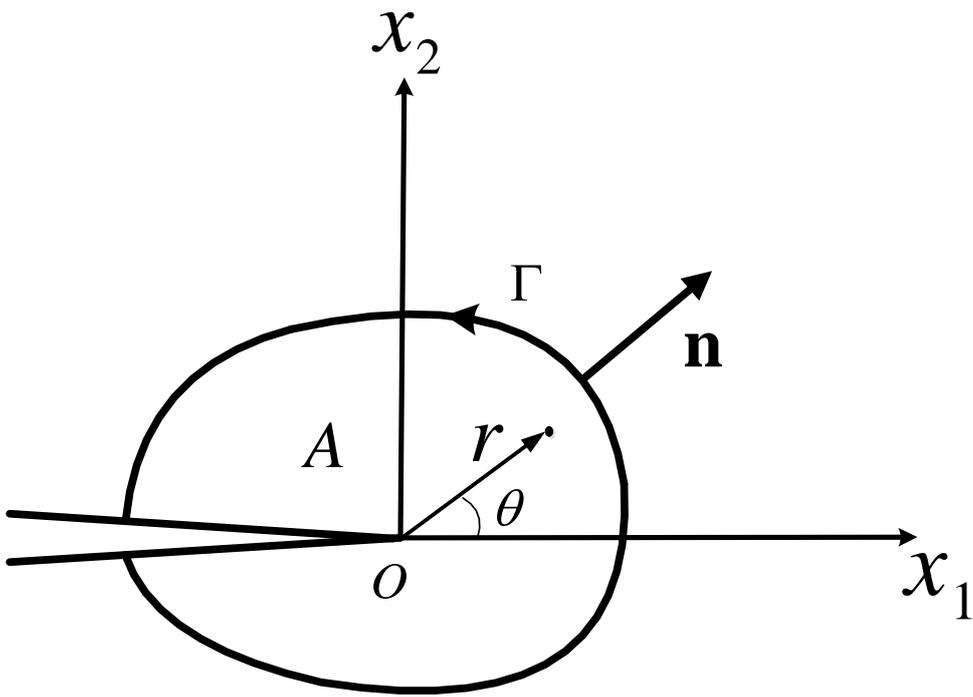

Fig. 1. Schematic of a stationary contour surrounding the crack tip.

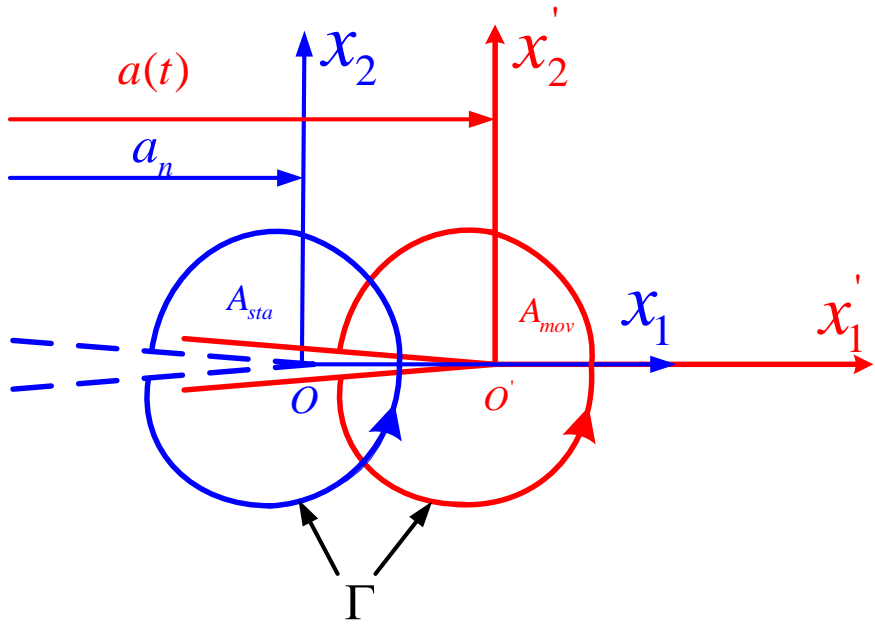

Fig. 2 Stationary and moving coordinate systems and corresponding contours.

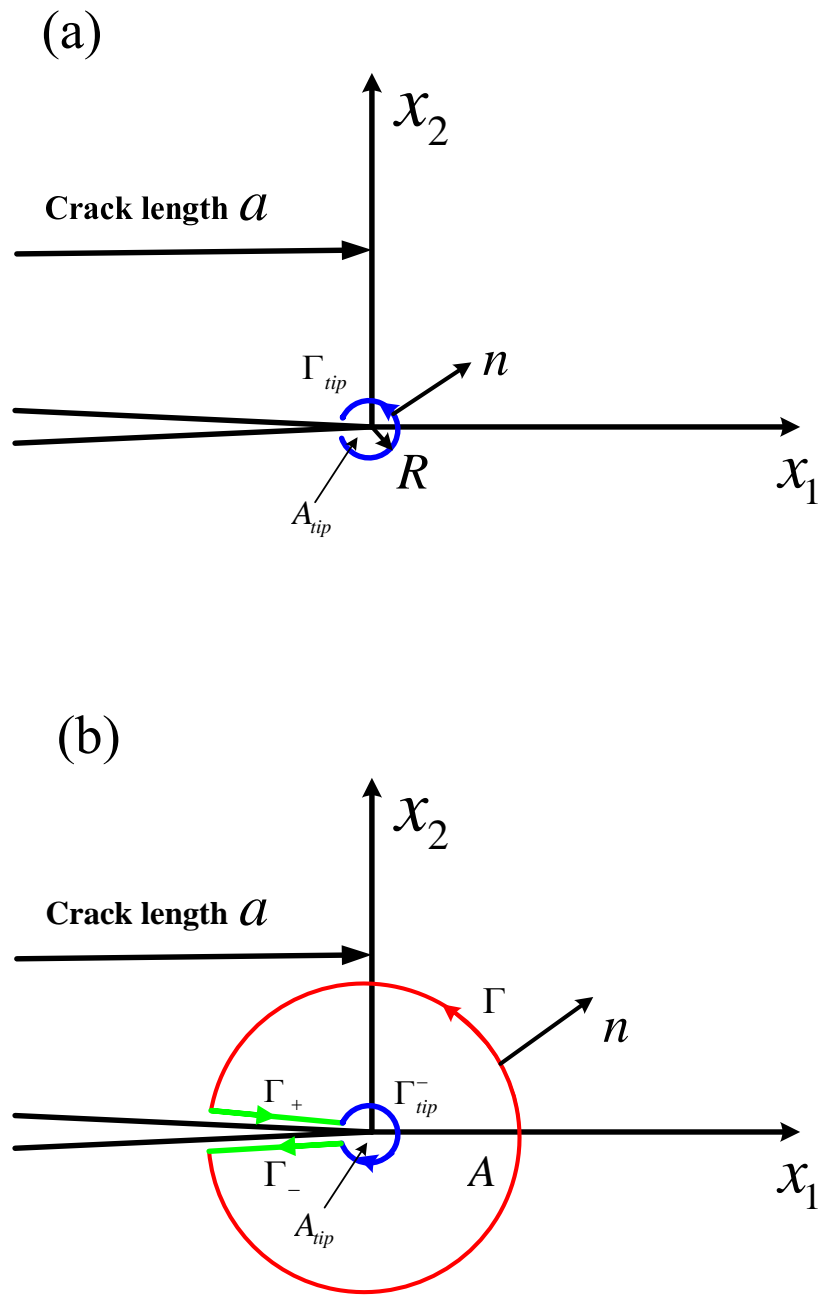

Fig. 3 Schematic diagrams of (a) an infinitesimal contour closely surrounding the crack tip and (b) an enclosed integration contour.

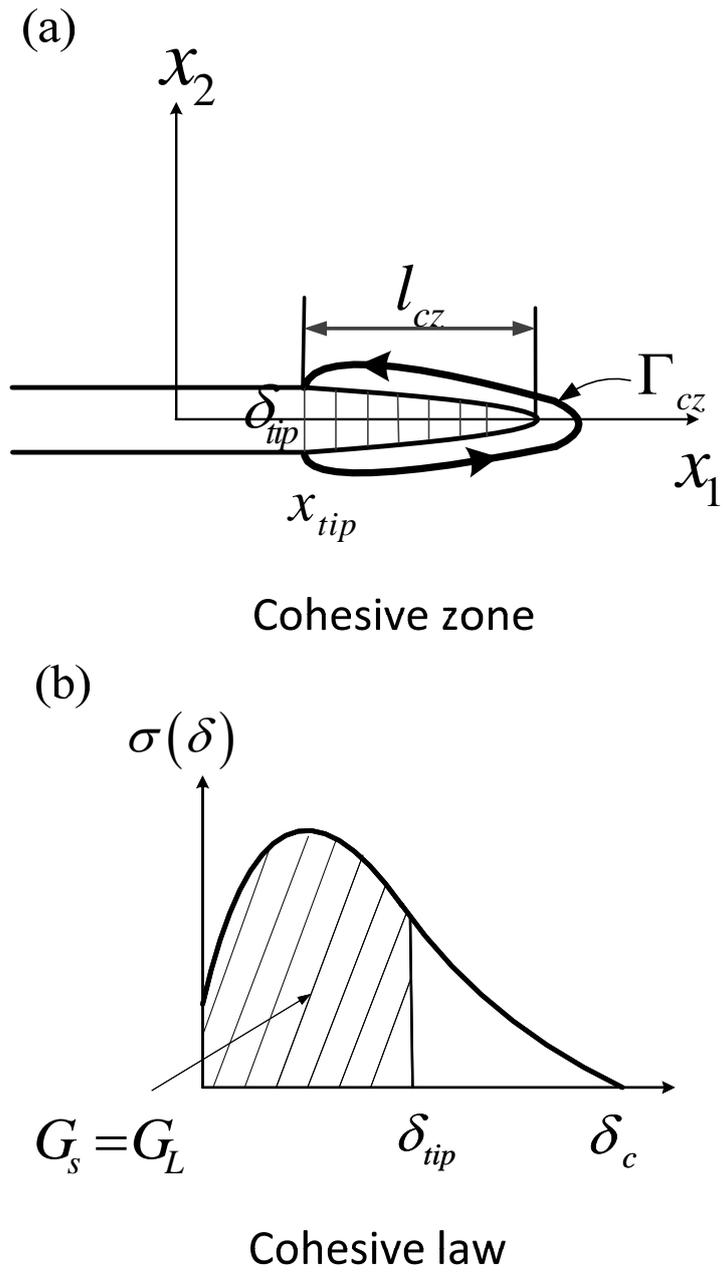

Fig. 4. Schematics of (a) a cohesive zone and (b) its cohesive law.

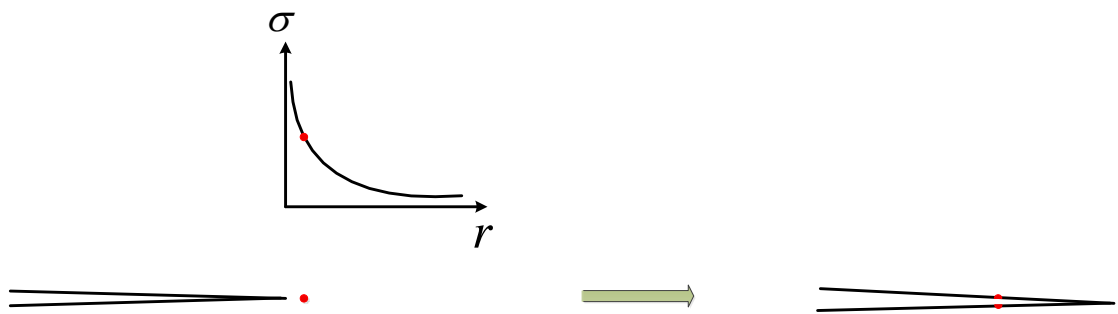

Fig. 5. Schematics of a material point on the crack plane before and after the crack passing through it with dramatic stress drop.

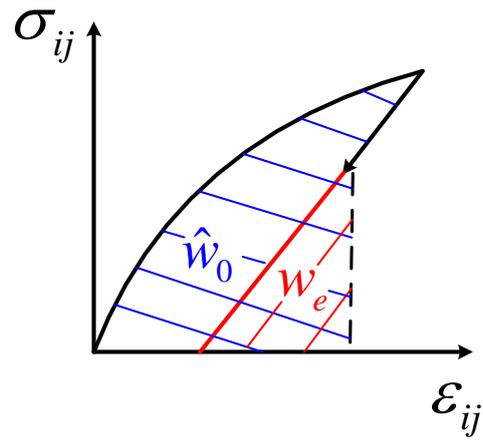

Fig. 6. Schematic of the elastic strain energy density.

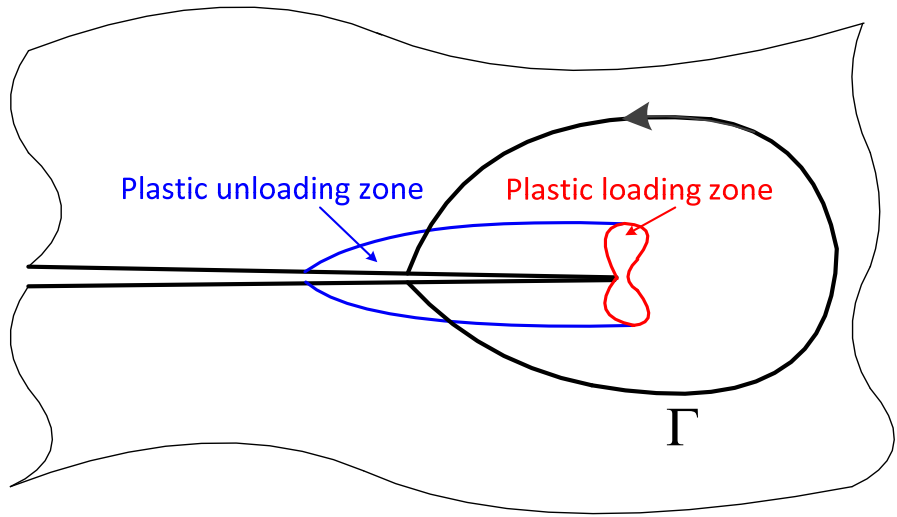

Fig. 7. Schematic of an integration contour enclosing the plastic loading zone for the global energy release rate $G$

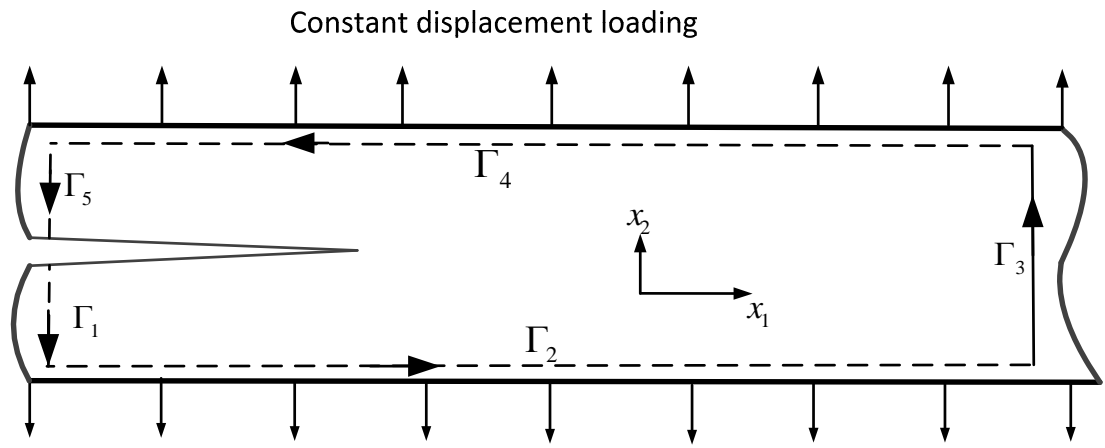

Fig. 8. A cracked strip with constant displacement loading on upper and lower boundaries and an integration contour.

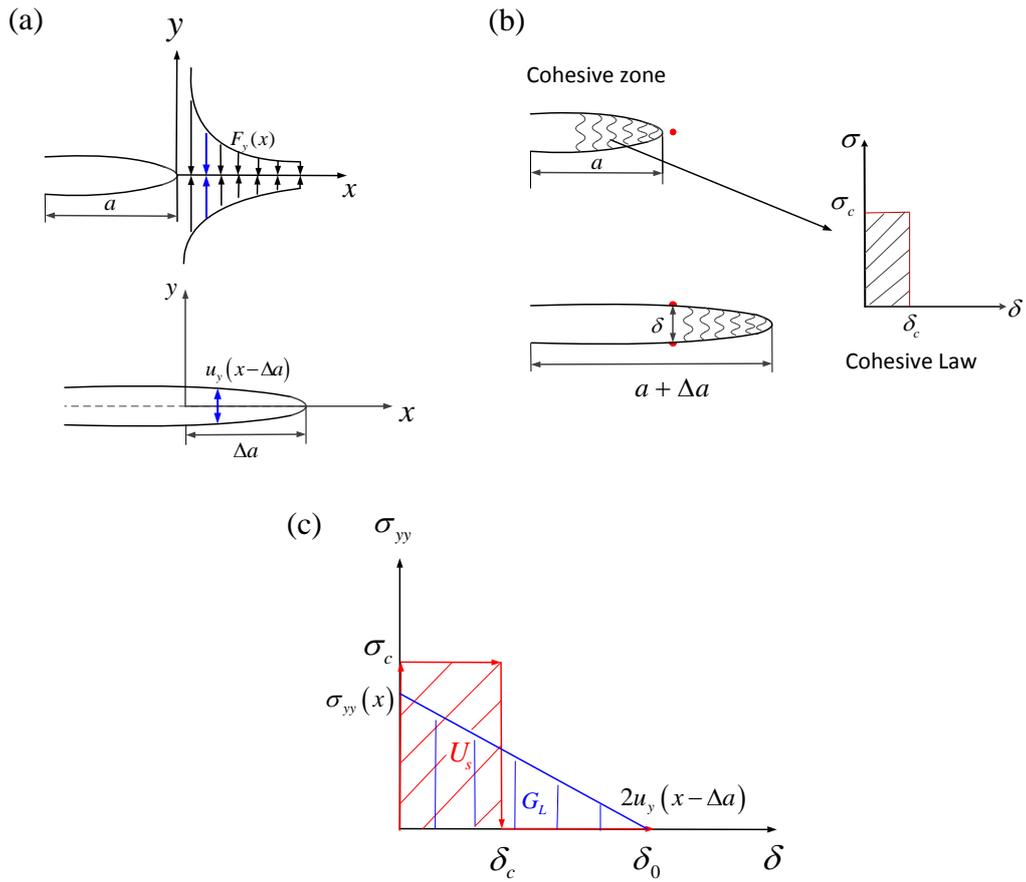

Fig. 9. Schematic snapshots of a propagating crack (a) in a linear elastic material and (b) in a material with a cohesive zone; (c) The corresponding works required.

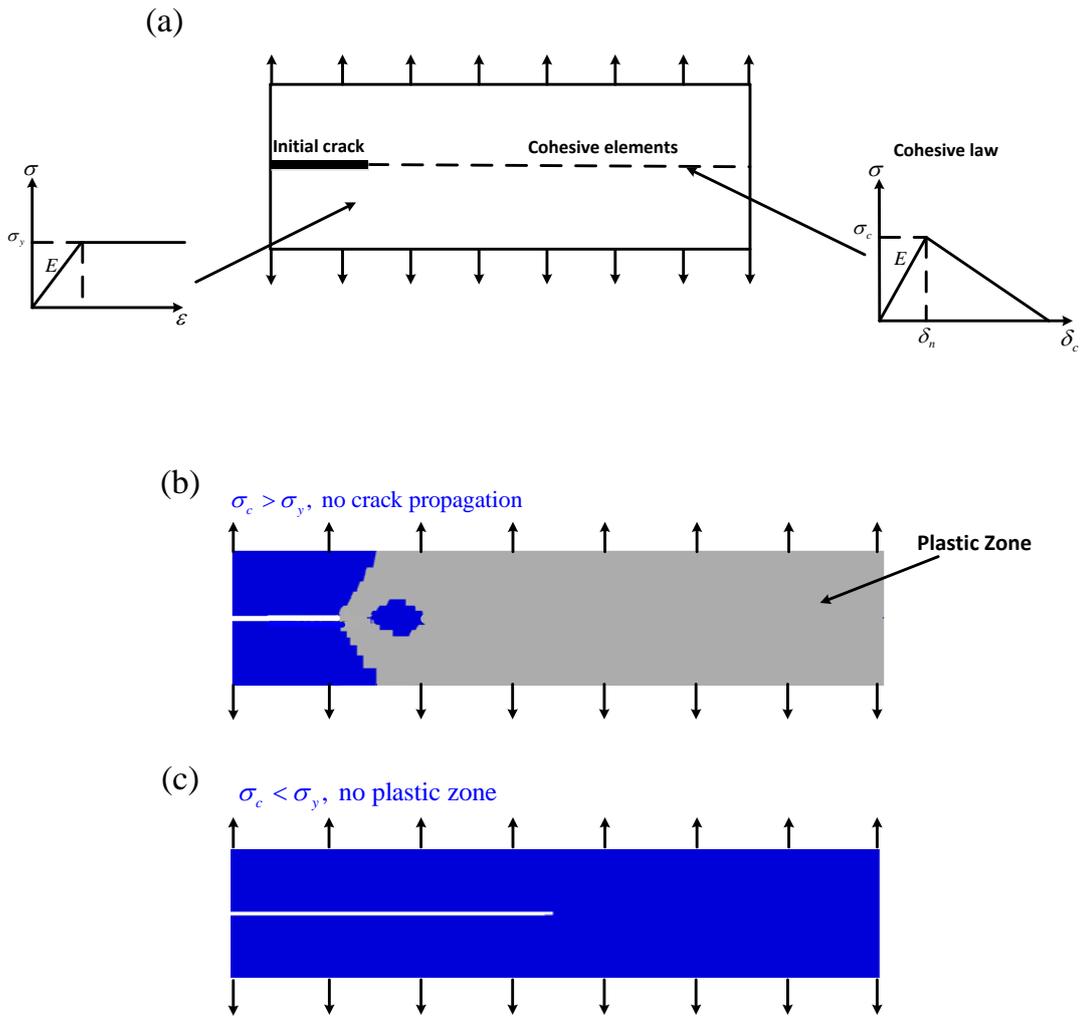

Fig. 10 (a) Schematic of a strip subject to constant displacement loading on the upper and lower boundaries with an initial crack and cohesive elements; (b) Simulation results of fracture for $\sigma_c > \sigma_y$ and (c) $\sigma_c < \sigma_y$.

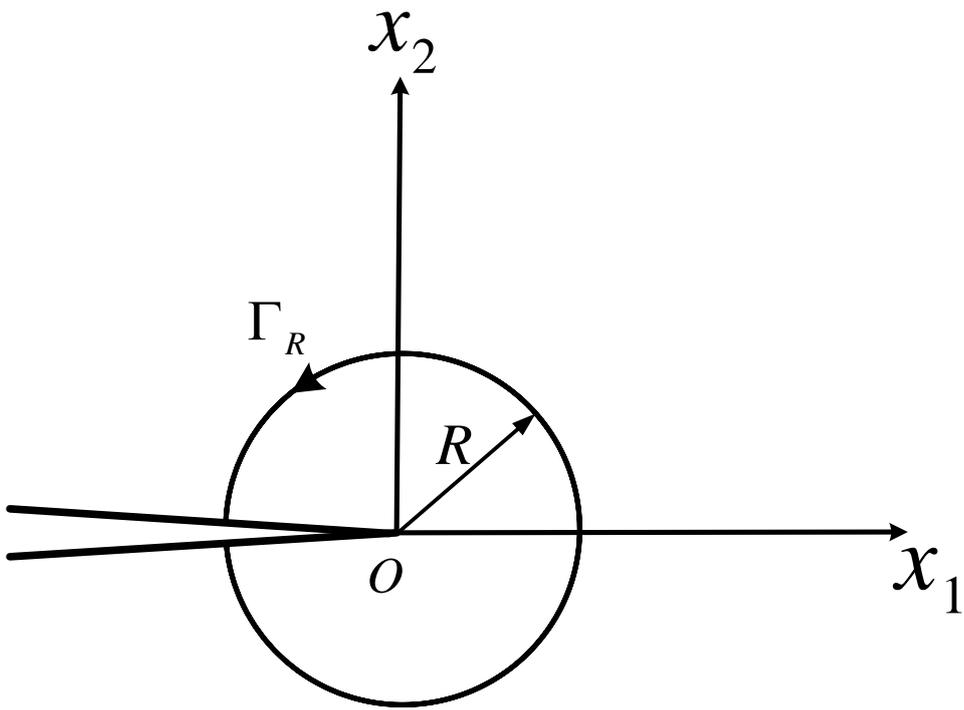

Fig. 11 Schematic of a finite circle contour

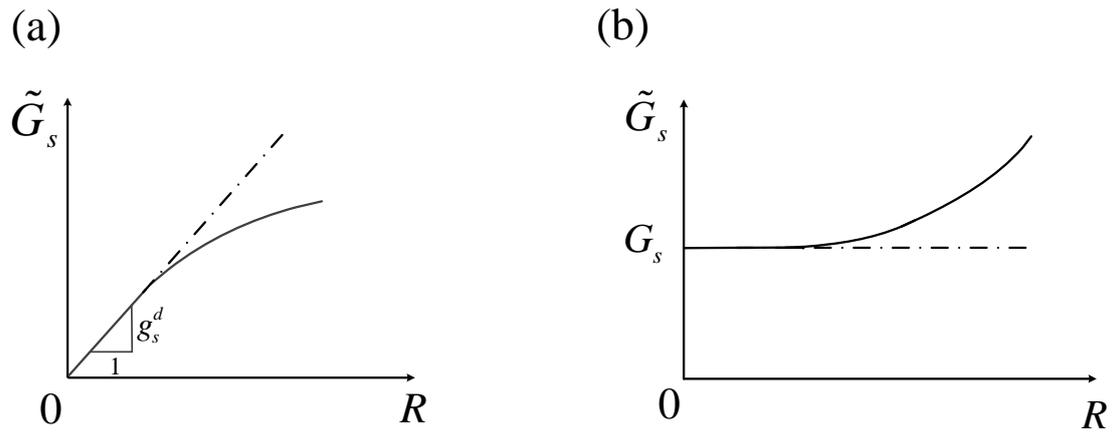

Fig. 12 Schematics of the relation between $\tilde{G}_s$ and $R$ for crack propagation in (a) elastic-perfectly plastic materials and (b) materials with a singular stress field at the tip.